\documentclass[aps,pra,amsmath]{revtex4}
\usepackage{graphicx}

\DeclareGraphicsExtensions{.eps}    

\newcommand{\Tr}{{\mathrm{Tr}}}
\newcommand{\psd}{\psi^{\dagger}}
\newcommand{\idt}{i\partial_t}
\newcommand{\ihdt}{i\hbar\partial_t}
\newcommand{\epo}{\varepsilon_{\pp}}
\newcommand{\varder[3]}{\frac{\delta #2}{\delta #3}}
\newcommand{\tph}{\widetilde{\varphi}}
\newcommand{\aos}{a_{\text{osc}}}
\newcommand{\nnc}{n_{\text{nc}}}
\newcommand{\bnc}{\tilde{n}_{\text{nc}}}
\newcommand{\bg}{\widetilde{g}}
\newcommand{\tdr}{\tilde{r}}
\newcommand{\rr}{\mathbf{r}}
\newcommand{\pp}{\mathbf{p}}
\begin{document}
\title{Effective action approach to a trapped Bose gas}
\author{Emil Lundh}
\altaffiliation[Present address:]{Helsinki Institute of Physics, P.\ O.\
Box 64, FIN-00014 University of Helsinki}
\author{J{\o}rgen Rammer}
\affiliation{Department of Theoretical Physics, Ume{\aa} University,
SE-901 87 Ume{\aa}}
\date{\today}
\begin{abstract}
The effective-action formalism is applied to a gas of
bosons. The equations describing the condensate
and the excitations are obtained using the loop expansion for the
effective action. For a
homogeneous gas the Beliaev expansion in terms of the diluteness
parameter is identified in terms of the loop expansion.
The loop expansion and the limits of validity of the well-known 
Bogoliubov and Popov
equations are examined analytically for a homogeneous dilute Bose gas and
numerically for a gas trapped in a harmonic-oscillator potential.
The expansion to one-loop order, and hence the Bogoliubov equation,
is shown to be valid for the zero-temperature
trapped gas as long as the characteristic length of the trapping
potential exceeds the s-wave scattering length.
\end{abstract}
\maketitle

\section{Introduction}

The dilute Bose gas has been subject to extensive study for
more than half a century, originally in an attempt to understand
liquid Helium II, but also as an interesting many-body system
in its own right. In 1947, Bogoliubov
showed how to describe Bose-Einstein condensation as a state of
broken symmetry, in which the expectation values of the
field operators are non-vanishing due to the single-particle
state of lowest energy being  macroscopically occupied, i.e.,
the annihilation and creation  operators for the
lowest-energy mode can be treated as c-numbers \cite{bogoliubov}.
In modern terminology, the expectation value of the
field operator is the order parameter and describes the density of the
condensed bosons.
In Bogoliubov's treatment, the physical quantities were expanded
in the diluteness parameter $\sqrt{n_0a^3}$, where $n_0$ denotes
the density of bosons occupying the lowest single-particle energy state,
and $a$ is the s-wave scattering length,
and Bogoliubov's theory is therefore only valid for homogeneous
dilute Bose gases. The inhomogeneous Bose gas was studied
by Gross \cite{gross} and Pitaevskii \cite{pitaevskii}, who
independently derived a nonlinear equation determining the
condensate density.
A field-theoretic diagrammatic treatment was applied by Beliaev
to the zero-temperature homogeneous dilute Bose gas, showing how
to go beyond Bogoliubov's approximation in a systematic
expansion in the diluteness parameter $\sqrt{n_0a^3}$
\cite{beliaev1,beliaev2}; and also showing
how repeated scattering 
leads to a renormalization of the interaction between the bosons.
This renormalization was in Beliaev's treatment a cumbersome
issue, where diagrams expressed in terms of
the propagator for the noninteracting
particles are intermixed with diagrams where the propagator contains 
the interaction
potential.
Beliaev's diagrammatic scheme was extended to finite temperatures by
Popov and Faddeev \cite{popovf}, and was subsequently employed to
extend the Bogoliubov theory to finite temperatures by incorporating
terms containing the excited-state operators to lowest order in the
interaction potential \cite{popov,popovbook}.

A surge of interest in the dilute Bose gas due to the experimental
creation of gaseous Bose-Einstein condensates occurred in the
mid-nineties \cite{jila}. The atomic condensates in the experiments
are confined in
external potentials, which poses new theoretical challenges;
especially, the Beliaev
expansion in the diluteness parameter $\sqrt{n_0a^3}$ is
questionable when the density is inhomogeneous. The
renormalization of the potential was generalized to a trapped
system by
Proukakis \textit{et al}.\ \cite{proukakis,proukakis2}.
Leading-order corrections to the
Gross-Pitaevskii equation for a trapped Bose gas were studied
by Stenholm \cite{stenholm}.
The finite temperature Beliaev-Popov theory was applied to a
trapped gas by Fedichev and Shlyapnikov \cite{fedichev}.

In this paper we shall employ the two-particle irreducible
effective-action approach,
and show that it provides an efficient systematic scheme for
dealing with both homogeneous Bose gases and
trapped Bose gases. We show how  the effective-action formalism
can be used to derive the equations
of motion for the dilute Bose gas, and more importantly that the loop
expansion can be used to determine
the limits of validity of approximations to the exact equations
of motion in the trapped case.

The paper is organized as follows. The model and the two-particle irreducible
effective-action approach are introduced
in Section \ref{formalismsec}. The
homogeneous dilute Bose gas is considered in Sec.\ \ref{homogensec}, and
familiar equations of motion are rederived.
In Sec.\ \ref{tmatrixsec}, we demonstrate how the renormalization of the
interaction potential due to repeated scattering is
conveniently carried out in the effective-action formalism. In Sec.\
\ref{trappedsec} we review the trapped Bose gas. The main results of the 
paper are presented in Sec.\ \ref{numericsec}, where
the equations of motion are solved numerically 
in order to assess 
the limits of validity of approximations to the exact equations of motion.
Finally, in Sec.\ \ref{conclusionsec} we summarize and conclude.

\section{Effective action formalism for Bosons}
\label{formalismsec}

We consider a system of spinless bosons described by the action
\begin{eqnarray}
  S[\psi,\psd] &=& \int d\rr dt
  \psd(\rr,t) \left[\idt-H_0(\rr)+\mu\right] \psi(\rr,t)
\nonumber\\&-&
  \frac12 \int d\rr d\rr' dt \psd(\rr,t)\psd(\rr',t)U(\rr-\rr')
  \psi(\rr',t)\psi(\rr,t)
\label{action}
\end{eqnarray}
where $\psi$ is the scalar field describing the bosons. Here
$\mu$ denotes the chemical potential, $H_0 = \pp^2/2m
+ V(\rr)$ is the one-particle Hamiltonian consisting of the kinetic
term and an external potential, and $U(\rr)$ is the
potential describing the interaction between the bosons.
We have chosen units so that $\hbar=1$, but
will restore $\hbar$ in final results.
It will prove convenient to introduce a matrix
notation whereby the field and its complex conjugate are combined  into
a two-component
field $\phi=(\psi,\psd)=(\phi_1,\phi_2)$.

The correlation functions of the Bose field are obtained from the
generating functional
\begin{equation}
\label{generating}
  Z[\eta,K] =
  \int {\cal D}\phi
  \exp \left( iS[\phi] + i\eta^{\dagger}
  \phi + \frac{i}{2} \phi^{\dagger} K \phi \right),
\end{equation}
by differentiating with respect to the source
$\eta^{\dag} = (\eta,\eta^{*})= (\eta_1,\eta_2)$. In
Eq.\ (\ref{generating}), matrix notation is implied in order to
suppress the integrations over space and time variables.
A two-particle source term, $K$, has been added to the action in the
generating functional
in order to obtain  equations involving the two-point
Green's function in a two-particle irreducible fashion.

The generator of the connected Green's functions is
\begin{equation}
  W[\eta,K]
  =
  -i \ln Z[\eta,K],
\end{equation}
and the derivative
\begin{equation}
  \frac{\delta W}{\delta \eta_i(\rr,t)}  =
  \bar{\phi}_i(\rr,t)
\end{equation}
gives the average
field, $\bar{\phi}$, with respect to the action
$S[\phi] + \eta^{\dag} \phi + \phi^{\dag} K \phi/2$,
\begin{equation}
\bar{\phi}(\rr,t)  =
\left(\begin{array}{ll}
\Phi(\rr,t)\\\Phi^*(\rr,t)
\end{array}\right) =
  \int {\cal D}\phi \,
  \phi(\rr,t)
  \exp \left( iS[\phi] + i\eta^{\dag}\phi +
\frac{i}{2} \phi^{\dag} K \phi \right)
=
\langle
{\phi}(\rr,t)
    \rangle .
\label{avefield}
\end{equation}
The average field $\Phi$ is seen to specify the
condensate density and is referred to as
the condensate wave function.

The derivative of $W$ with respect to the two-particle source is
\begin{eqnarray}
  \frac{\delta W}{\delta K_{ij}(\rr,t;\rr',t')}
& = &
  \frac{1}{2}\bar{\phi}_i(\rr,t)
  \,\bar{\phi}_{j}(\rr',t')
  \nonumber\\
  & + &
  \frac{i}{2} G_{ij}(\rr,t,\rr',t'),
\end{eqnarray}
where $G$ is the full connected two-point matrix Green's function
describing the bosons not in the condensate,
\begin{eqnarray}
  G_{ij}(\rr,t,\rr',t')
   =
  - \frac{\delta^2 W}{\delta \eta_i(\rr,t) \, \delta \eta_j(\rr',t')}
  =
  -i \left(
  \begin{array}[c]{cc}
    \langle \delta\psi(\rr,t) \delta\psd(\rr',t')
 \rangle
 & \langle \delta\psi(\rr,t) \delta\psi(\rr',t')
    \rangle\\
    \langle
\delta\psd(\rr,t) \delta\psd(\rr',t')
\rangle
    & \langle
\delta\psd(\rr,t) \delta\psi(\rr',t')
   \rangle
  \end{array}
 \right),
\label{greensdef}
\end{eqnarray}
where $\delta\psi(\rr,t)$ is the deviation of the field from its
mean value, $\delta\psi = \psi - \Phi$. Likewise, we shall write
$\phi = \bar{\phi} + \delta \phi$ for the two-component
field.
We note that in the path integral representation, averages over
fields, such as in Eq.\ (\ref{greensdef}),
are automatically
time-ordered.

We introduce the effective
action, $ \Gamma $, the generator of the two-particle irreducible vertex
functions, through the Legendre transform of the generator of connected
Green's functions, $W$:
  \begin{eqnarray}
    \Gamma[\bar{\phi},G] & = & W[\eta,K] -
    \eta^{\dag} \bar{\phi} -
    \frac{1}{2} \bar{\phi}^{\dag} K \bar{\phi} -
    \frac{i}{2} \Tr GK.
\end{eqnarray}
The effective action satisfies the equations
$\delta \Gamma/\delta \bar{\phi} = -\eta - K \bar{\phi}$
and  $\delta \Gamma/\delta G = - i K/2$.
In a physical state where the external sources vanish, $\eta=0=K$,
the variations of the effective action with respect to the
field averages $\bar{\phi}$ and $G$ vanish, yielding the
equations of motion:
\begin{subequations}
\begin{eqnarray}
    \label{eq:motion1}
  \frac{\delta \Gamma}{\delta \bar{\phi}} = 0,
\\
\label{eq:motion2}
  \frac{\delta \Gamma}{\delta G} = 0. 
\end{eqnarray}
\end{subequations}

According to Cornwall {\em et al.} \cite{cornwall}, the effective
action can be written on the form
\begin{equation}\label{effaction}
  \Gamma[\bar{\phi},G] = S[\bar{\phi}] +
  \frac{i}{2} \Tr \ln G_0 G^{-1} +
  \frac{i}{2} \Tr (G_0^{-1}-\Sigma^{(1)})G
 - \frac{i}{2}\Tr 1+
\Gamma_2[\bar{\phi},G],
\end{equation}
where $G_0$ is the noninteracting matrix Green's function,
\begin{equation}
  G_0^{-1}(\rr,t,\rr',t') = -\left(\begin{array}{ll}
  \idt-H_0+\mu & 0\\
  0 & -\idt-H_0+\mu
  \end{array}\right)\delta(\rr-\rr')\delta(t-t'),
\end{equation}
and the matrix
\begin{equation}
  \Sigma^{(1)}(\rr,t,\rr',t') =
  - \left.
  \frac{\delta^2 S}{\delta\phi^{\dagger}(\rr,t)\delta \phi(\rr',t')}
  \right|_{\phi=\bar{\phi}} + G_0^{-1}(\rr,t,\rr',t'),
\end{equation}
will turn out to be the self-energy to one-loop order
(see Eq.\ (\ref{dysonbel})). Using the action describing the
bosons, Eq.\ (\ref{action}),
we obtain for the components
$\Sigma^{(1)}_{ij}(\rr,t,\rr',t') =
\delta(t-t')\Sigma^{(1)}_{ij}(\rr,\rr')$, where
\begin{eqnarray}
  \Sigma^{(1)}_{11}(\rr,\rr')&=& \delta(\rr'-\rr)
  \int d\rr'' U(\rr-\rr'')|\Phi(\rr'',t)|^2
  + U(\rr-\rr')\Phi^*(\rr',t)\Phi(\rr,t),\nonumber\\
  \Sigma^{(1)}_{12}(\rr,\rr')&=& U(\rr-\rr')
  \Phi(\rr,t)\Phi(\rr',t), \nonumber\\
  \Sigma^{(1)}_{21}(\rr,\rr')&=& U(\rr-\rr')
  \Phi^*(\rr,t)\Phi^*(\rr',t),\nonumber\\
  \Sigma^{(1)}_{22}(\rr,\rr')&=& \delta(\rr'-\rr)
  \int d\rr'' U(\rr-\rr'')|\Phi(\rr'',t)|^2
  + U(\rr-\rr')\Phi^*(\rr,t)\Phi(\rr',t).
\label{sigma0}
\end{eqnarray}
The delta function in the time coordinates reflects the fact that the
interaction is instantaneous.
Finally, the quantity $\Gamma_2$ in Eq.\ (\ref{effaction})
is
\begin{eqnarray}
\Gamma_2 = -i \ln \langle e^{i
  S_{\text{int}}[\bar{\phi},\delta\phi]}
\rangle_G^{2\rm PI},
\end{eqnarray}
where
$S_{\rm int}[\bar{\phi},\delta\phi]$ denotes the part of the action
$S[\bar{\phi}+\delta\phi]$
which is higher than second order in $\delta\phi$ in an expansion around the
average field.
The quantity
$\Gamma_2 $ is conveniently described in terms of the diagrams
generated by the action $S_{\rm int}[\bar{\phi},\delta\phi]$, and
consists of all the two-particle irreducible vacuum diagrams
as indicated by
the superscript ``2PI'', and
the diagrams will therefore contain two or more loops.
The subscript indicates that propagator lines represent the full Green's
function $G$, i.e.,
the brackets with subscript $G$ denote the average
\begin{equation}
  \langle e^{i S_{\rm int}[\bar{\phi},\delta \phi]} \rangle_G
  =
  (\det iG)^{-1/2} \int {\cal D}(\delta \phi)
  e^{\frac{i}{2} \delta \phi^{\dag} G^{-1} \delta \phi }
  e^{i S_{\rm int}[\bar{\phi},\delta \phi]}.
\end{equation}

\begin{figure}
\includegraphics[width=\columnwidth]{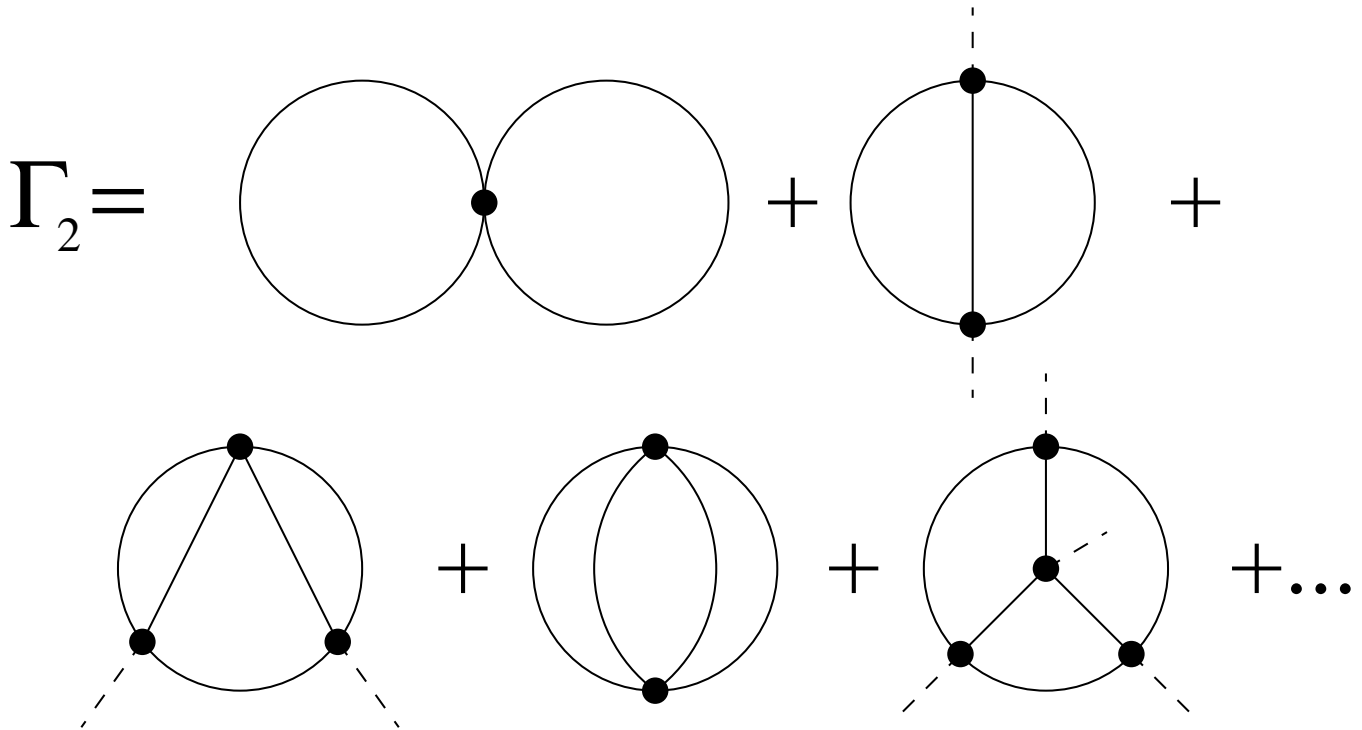}
\caption{\label{twoloopfig}Two-loop (upper row) and three-loop
vacuum diagrams (lower row) contributing to the effective action.}
\end{figure}

The diagrammatic expansion of $\Gamma_2$ corresponding to the action
for the bosons, Eq.\ (\ref{action}), is
illustrated  in Figure \ref{twoloopfig} where the two-
and three-loop vacuum diagrams are shown.
Since matrix indices are suppressed the diagrams are
to be understood as follows. Full lines represent Green's
functions and in the cases where we display the different components
explicitly, $G_{11}$ will carry one arrow ($G_{22}$ can according to
Eq.\ (\ref{greensdef}) be expressed in terms of $G_{11}$ and thus needs no
special symbol), $G_{12}$ has
two arrows pointing inward and $G_{21}$ carries two arrows pointing
outward. Dashed lines represent the condensate wave function and can
also be decorated with arrows, directed out from
the vertex to represent $\Phi$, or directed towards the vertex
representing $\Phi^*$. The dots where four lines meet are
interaction vertices, i.e., they represent the interaction
potential $U$ (which in other contexts will be represented by
a wiggly line). When
all possibilities for the indices are exhausted, subject to the
condition that each vertex has two ingoing and two outgoing particle
lines, we have
represented all the terms of $\Gamma_2$ to a given loop order.
Finally, the expression corresponding to
each vacuum diagram should be multiplied by the factor
$i^{s-2}$, where $s$ is the number of loops the diagram contains.

It is well-known that the expansion of the effective action
in loop orders is an expansion in
Planck's constant $\hbar$ \cite{cornwall}.
The first term $S[\bar{\phi}]$
on the right-hand side of Eq.\ (\ref{effaction}) is referred to
as the zero-loop term
and the terms where the trace is written explicitly as one-loop terms,
because they are
proportional to $\hbar^0$ and $\hbar^1$, respectively.
We note that the presented effective action approach is capable of
describing arbitrary
states, including non-equilibrium situations where the external potential
depends on time. Although we in the present paper shall limit
ourselves to study a Bose gas at zero temperature the theory is
straightforwardly generalized to finite temperatures. For example,
in the Schwinger-Keldysh technique for treating general
  non-equilibrium states, the field is just attributed an additional
  index. For an account of treating
arbitrary states we refer to
Ref.\ \cite{GR}; see also Ref. \ref{stoofkeldysh} for an application 
of the Schwinger-Keldysh technique to the dilute Bose gas.
The equations of motion
(\ref{eq:motion1}-\ref{eq:motion2})
 together with the expression for the effective action, Eq.\ (\ref{effaction}),
form the basis for our
subsequent calculations.


\section{Homogeneous Bose gas}
\label{homogensec}

We shall now consider the case of a homogeneous Bose gas in equilibrium. 
The theory of such a system is well known, but the effective action 
formalism will prove to be a simple and efficient tool which permits one to 
derive the well-known results with particular ease.
For the case of a
homogeneous Bose gas in equilibrium, the general theory presented in
the previous section simplifies considerably.
The single-particle  Hamiltonian $H_0$ is then simply equal to
the kinetic term,
$H_0(\pp)=\pp^2/2m\equiv \epo$,
and the condensate wave function $\Phi(\rr,t)$ is a
time and coordinate-independent constant whose value is denoted by
$\sqrt{n_0}$, so that $n_0$ denotes the condensate density.
The first term in the effective action, Eq.\ (\ref{effaction}), is
then
\begin{equation}
  S[\Phi] = (\mu n_0 -\frac12 U_0 n_0^2) \int d\rr dt 1,
\end{equation}
where $U_0 = \int d\rr U(\rr)$ is the zero-momentum component of the
interaction potential. For a constant value of the
condensate wave function, $\Phi(\rr,t)=\sqrt{n_0}$, Eq.\
(\ref{sigma0}) yields
\begin{equation}
  \Sigma^{(1)}(\pp) = \left(\begin{array}{ll}
  n_0(U_0+U_{\pp}) & n_0U_{\pp}\\
  n_0U_{\pp} & n_0(U_0+U_{\pp})
  \end{array}\right).
\label{sigma12001}
\end{equation}
Varying, in accordance with Eq.\ (\ref{eq:motion1}),
the effective action, Eq.\ (\ref{effaction}), with
respect to $n_0$
yields the equation for
the chemical potential
\begin{eqnarray}
  \mu &=& n_0 U_0 +\frac{i}2\int \frac{d^4p}{(2\pi)^4}
  \left[(U_0+U_{\pp})(G_{11}(p)+G_{22}(p))+
  U_{\pp}(G_{12}(p)+G_{21}(p))
  \right] -
  \varder{\Gamma_2}{n_0},
\label{mueq2001}
\end{eqnarray}
where the notation for the four-momentum, $p=(\pp,\omega)$,
has been introduced.
The first term on the right-hand side is the zero-loop result, which
depends only on the condensate fraction of the bosons. The second term on the
right-hand side is the one-loop term which takes the noncondensate
into account. The term involving the anomalous Green's functions
$G_{12}$ and $G_{21}$ will shortly be absorbed by the renormalization
of the interaction potential (see Sec.\ \ref{tmatrixsec}).
From the last term originate the higher loop terms
which will be dealt with at the end of the section.

The equation determining the Green's function
is obtained by varying the
effective action with respect to the matrix Green's function
$G(p)$, in accordance with Eq.\ (\ref{eq:motion2}), yielding
\begin{equation}
  0 = \varder{\Gamma}{G} =
  -\frac{i}2\left(-G^{-1}+G_0^{-1}+ \Sigma^{(1)} +\Sigma'\right),
\label{dysonbel1}
\end{equation}
where
\begin{equation}
\Sigma'_{ij}=2i
\frac{
\delta\Gamma_2
}{
\delta G_{ji}
}
.
\label{sigmafraGamma}
\end{equation}
Introducing the
notation for
the matrix self-energy $\Sigma = \Sigma^{(1)}+\Sigma'$,
Eq.\ (\ref{dysonbel1}) is seen to be the Dyson
equation:
\begin{equation}
  G^{-1} = G_0^{-1} - \Sigma .
  \label{dysonbel}
\end{equation}
In the
context of the dilute Bose gas, this equation is referred to as
the Dyson-Beliaev equation.

The Green's function in
momentum space is obtained by simply inverting the 2$\times$2 matrix
$G_0^{-1}(p)-\Sigma(p)$ resulting in the following components:
\begin{eqnarray}
  G_{11}(p) &=& \frac{\omega+\epo-\mu+\Sigma_{22}(p)}{D_p},\nonumber\\
  G_{12}(p) &=& \frac{-\Sigma_{12}(p)}{D_p},\nonumber\\
  G_{21}(p) &=& \frac{-\Sigma_{21}(p)}{D_p},\nonumber\\
  G_{22}(p) &=& \frac{-\omega+\epo-\mu+\Sigma_{11}(p)}{D_p},
\label{momentumgreens}
\end{eqnarray}
all having the common denominator
\begin{equation}
  D_p = (\omega+\epo-\mu+\Sigma_{22}(p))(\omega-\epo+\mu-\Sigma_{11}(p))
  +\Sigma_{12}(p)\Sigma_{21}(p).
\end{equation}
From the expression for the matrix Green's function, Eq.\
(\ref{greensdef}), it follows that in the
homogeneous case its components obey the relationships
$G_{22}(p)=G_{11}(-p)$ and
$G_{12}(-p)=G_{12}(p)=G_{21}(p)$. The
corresponding relations hold for the self-energy components.
We note that the results found for $\mu$ and $G$ to zero- and one-loop
order coincide with those found in Ref.\ \cite{beliaev1}
to zeroth and first order in the diluteness parameter
$\sqrt{n_0a^3}$. For example, according to
Eq.\ (\ref{sigma12001}) we obtain for the components of the
matrix Green's function to one loop order
\begin{eqnarray}
G_{11}^{(1)}(p) =
\frac{\omega+\epo + n_0
U_{\pp}}{\omega^2-\epo^2-2n_0U_{\pp} \epo},
\nonumber\\
G_{12}^{(1)}(p) =
\frac{-n_0U_{\pp}}{\omega^2-\epo^2-2n_0U_{\pp} \epo},
\label{zerogreens}
\end{eqnarray}
which are the same expressions as the ones in Ref.\ \cite{beliaev1}.
As we shortly demonstrate, the loop expansion for the
case of a homogeneous Bose gas is in fact
equivalent to an expansion in the diluteness parameter.
From Eq.\ (\ref{zerogreens}) we obtain for the single-particle
excitation energies to one-loop order
$E_{\pp} = \sqrt{\epo^2+2n_0U_{\pp} \epo}$, which are the well-known
Bogoliubov energies \cite{bogoliubov}.

Differentiating with respect to
$n_0$ the terms in $\Gamma_2$ corresponding to the
two-loop vacuum diagrams 
gives the two-loop contribution to the chemical potential.
Functionally differentiating the same terms with respect to
$G_{ji}$ gives
the two-loop contributions to the self-energies $\Sigma_{ij}$.
The diagrams we thus obtain for the chemical potential $\mu$ and the
self-energy $\Sigma$ are topologically identical to
those found by Beliaev
\cite{beliaev2};
however, the interpretation differs in that the propagator in the
vacuum diagrams of Fig.\ \ref{twoloopfig} is the exact
propagator, whereas in reference \cite{beliaev2} the propagator to
one-loop order appears.

In order to establish that the loop expansion for a homogeneous Bose
gas is an expansion in the
diluteness parameter $\sqrt{n_0a^3}$, we examine the general structure
of the vacuum diagrams comprised by $\Gamma_2$.
Any diagram of a given loop
order differs from any diagram in the preceding loop order by
an extra four-momentum integration, the condensate density $n_0$
to some power $k$, the interaction potential $U$ to the power
$k+1$, and $k+2$ additional Green's functions in the integrand.
We can estimate the contribution from these terms as follows.
The Green's functions are approximated by the one-loop result
Eq.\ (\ref{zerogreens}). The additional frequency
integration over a product of $k+2$ Green's functions
 yields $k+2$ factors
of $n_0U$ (where $U$ denotes the typical magnitude of the
Fourier transform of the interaction potential),
divided by $2k+3$ factors of the Bogoliubov energy $E$.
The range of the momentum integration provided
by the Green's functions is $(mn_0U)^{1/2}$. The remaining
three-momentum integration therefore gives
a factor of order $n_0^{-k+1/2} m^{3/2} U^{-k+1/2}$,
and provided the Green's functions make the integral converge,
the contribution from an additional loop is of the order
$(n_0m^3U^3)^{1/2}$.
This is the case except for the so-called ladder
diagrams, in which case the convergence need to be provided by the momentum
dependence of the potential.
The ladder diagrams will be dealt with separately in the
next section where we show that they through a renormalization of the
interaction potential lead to the appearance of the t-matrix which in the
dilute limit is proportional to the s-wave scattering length $a$ and
inversely proportional to the boson mass.
The renormalization of the interaction
potential will therefore
not change the estimates performed above, but only change the
expansion parameter. Anticipating this change we conclude
that the expansion parameter governing the loop expansion is
for a homogeneous Bose gas indeed identical
to Bogoliubov's diluteness parameter $\sqrt{n_0a^3}$.

\section{Renormalization of the interaction}
\label{tmatrixsec}

\begin{figure}[htbp]
\includegraphics[width=\columnwidth]{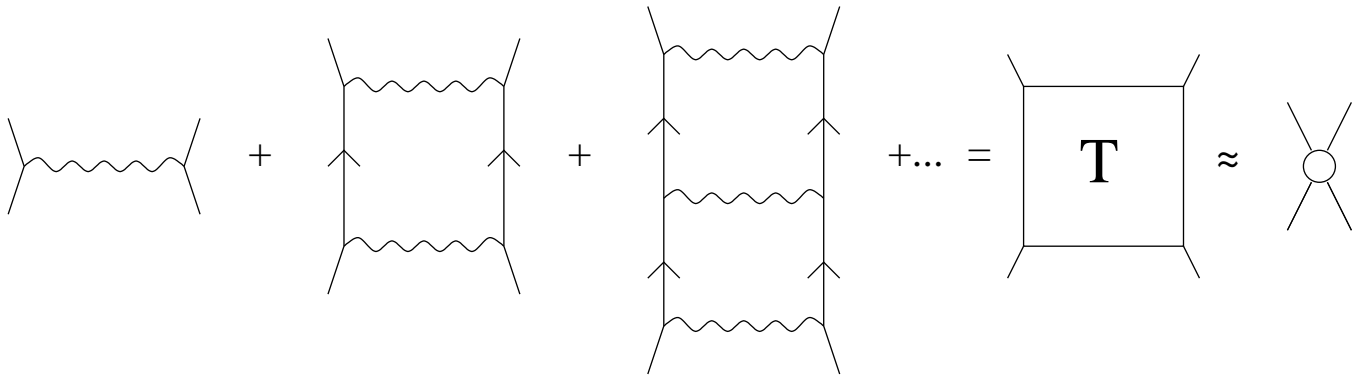}
\caption{\label{tmatrixfig}Summing all diagrams of the
``ladder'' type results in the t-matrix, which to lowest order in the
diluteness parameter is a momentum independent constant $g$, 
diagrammatically represented by a circle.
}
\end{figure}

Instead of having the interaction potential
appear explicitly in diagrams, one should work in
the skeleton diagrammatic representation where diagrams are summed so
that the four-point vertex
appears instead of the interaction potential, thus accounting for the
repeated scattering of the bosons.
In the dilute limit, where the inter-particle distance is large
compared to the s-wave scattering length,
the so-called ladder diagrams give the
largest contribution to the four-point vertex function \cite{griffinreview}.
The ladder diagrams are depicted in Fig.\ \ref{tmatrixfig}.
On computing the corresponding integrals, it is found
that an extra ``rung'' in a ladder contributes with a factor
proportional not to $\sqrt{n_0m^3U^3}$ as was the case for the type of
extra loops considered in the previous section, but to
$k_0mU$, where $k_0$ is the upper cut-off momentum (or
inverse spatial range) of the potential, as first noted by Beliaev 
\cite{beliaev2}. The quantity $k_0mU$ is not
necessarily small for the atomic gases under consideration here.
Hence, all vacuum diagrams which differ only in the number of ladder rungs
that they contain are of the same order in the diluteness parameter,
and we have to perform a summation over this infinite class of
diagrams.
The ladder resummation
results in an effective potential $T(p,p',q)$, which is called
the t-matrix and is a function of the two ingoing momenta and
the four-momentum transfer. Due to the instantaneous nature of
the interactions, the t-matrix does not depend on the frequency
components of the ingoing four-momenta, but for notational convenience
we display the dependence as $T(p,p',q)$.
To lowest order in the diluteness parameter, 
the t-matrix is independent of four-momenta and proportional to
the constant scattering amplitude, $T(0,0,0) = 4\pi\hbar^2 a/m = g$,
where $a$ is the s-wave scattering length \cite{griffinreview,stoofnist}.
This is illustrated
in Fig.\ \ref{tmatrixfig}, where we have chosen an open circle to
represent $g$.
Iterating the equation for the ladder diagrams we obtain the well-known
t-matrix equation:
\begin{equation}
T(p,p',q) = U_{\mathbf{q}} +
i \int d^4q' U_{\mathbf{q}'} G_{11}(p+q') G_{11}(p-q')
T(p+q', p-q', q-q').
\end{equation}
If the theory is generalized to finite temperatures, the t-matrix
takes into account the effects of thermal population of the excited states.

\begin{figure}[htbp]
\includegraphics[width=\columnwidth]{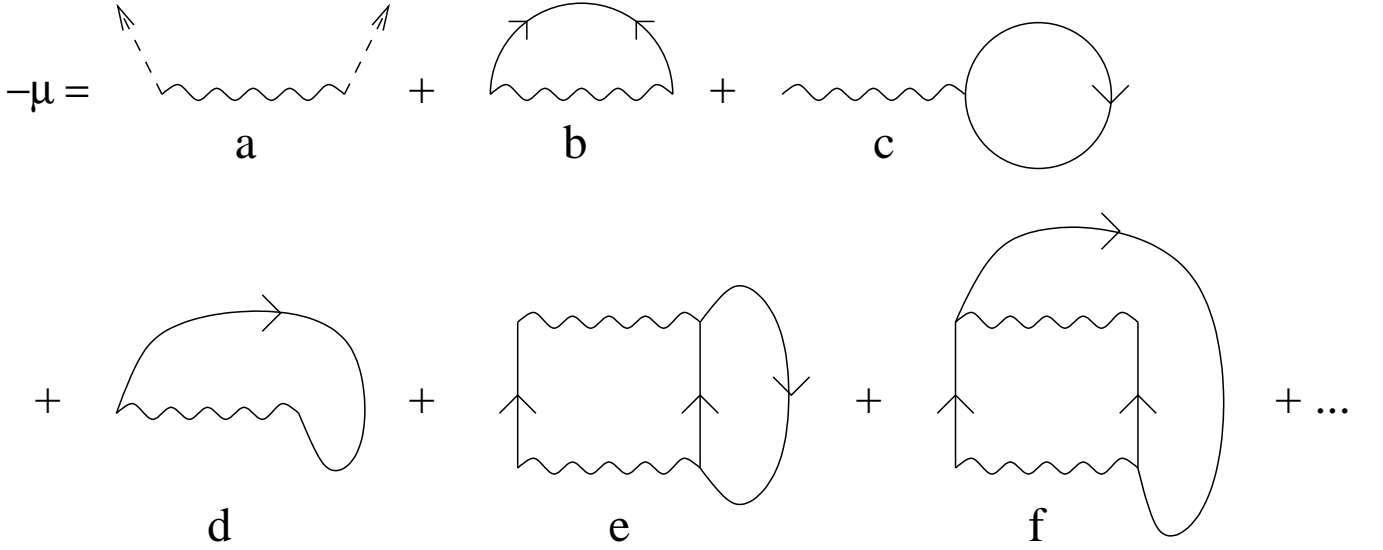}
\caption{\label{mufig}Diagrams up to two-loop order
contributing to the chemical potential. Only
the two-loop diagrams relevant to the resummation of
the ladder diagrams are displayed. The two-loop diagrams not displayed are
topologically identical to those shown, but differ in the direction of
arrows or the presence of anomalous instead of normal propagators.}
\end{figure}

We shall now show how the ladder resummation alters the diagrammatic
representation of the chemical potential and the self-energy (see also 
Ref.\ \cite{griffinreview}). In Fig.\
\ref{mufig} is displayed some of the terms up to two-loop order contributing
to the chemical potential $\mu$. The first two terms in Eq.\ (\ref{mueq2001})
is represented by diagrams  a-d, and the two-loop diagrams e-f
originate from $\Gamma_2$. The diagrams labeled e and
f are formally one loop order higher than c and d, but they differ
only by containing one additional ladder rung.
Hence, the diagrams c, d, e, and f, and all the diagrams
that can be constructed from these by adding ladder rungs, are of
the same order in the diluteness parameter $\sqrt{n_0a^3}$
as showed at the end of the previous section. They are therefore
resummed, and as discussed this leads to the replacement of the
interaction potential $U$ by the t-matrix.

\begin{figure}[htbp]
\includegraphics[width=\columnwidth]{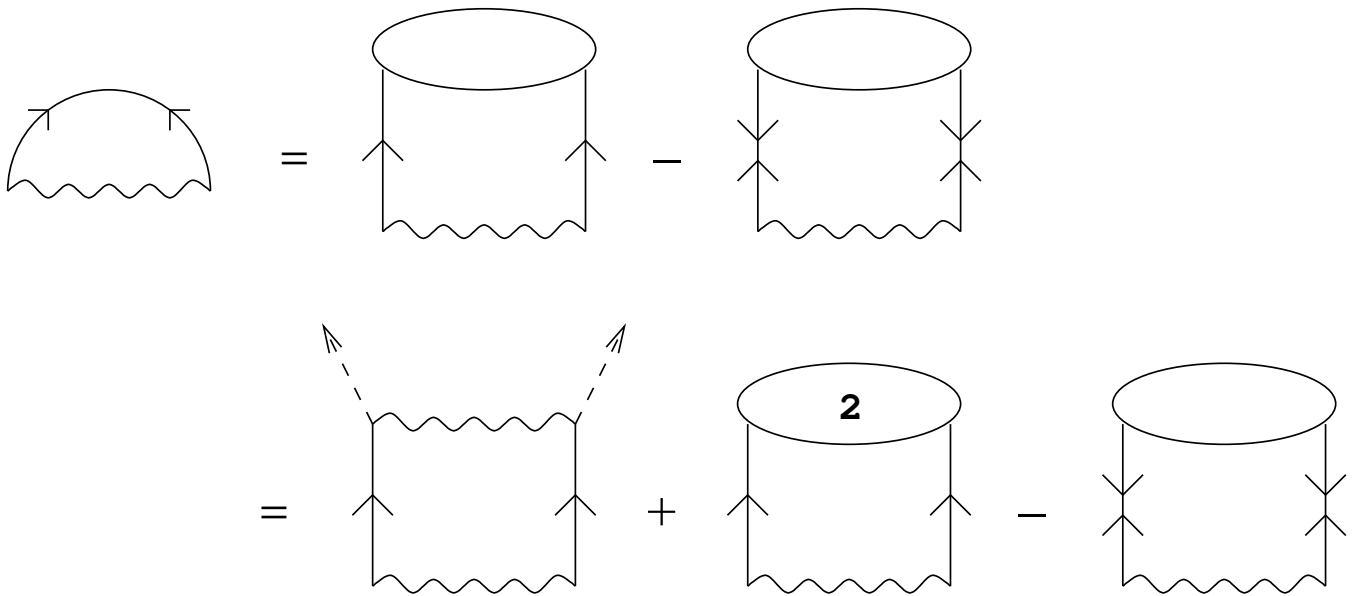}
\caption{\label{laddermanipfig}Diagrammatic representation of the
last two rewritings in  Eq.\ (\ref{laddermanip})
which lead to the conclusion that the diagram b of Fig.\
\ref{mufig} implicitly contains the ladder contribution to diagram
a. The anomalous self-energy $\Sigma_{12}$ is represented by an oval
with two ingoing lines, $\Sigma_{21}$ is represented by an oval
with two outgoing lines, and the sum of the second and higher-order
contributions to $\Sigma_{12}$ is represented by an oval with the
label ``2''.}
\end{figure}

\begin{figure}[htbp]
\includegraphics[width=\columnwidth]{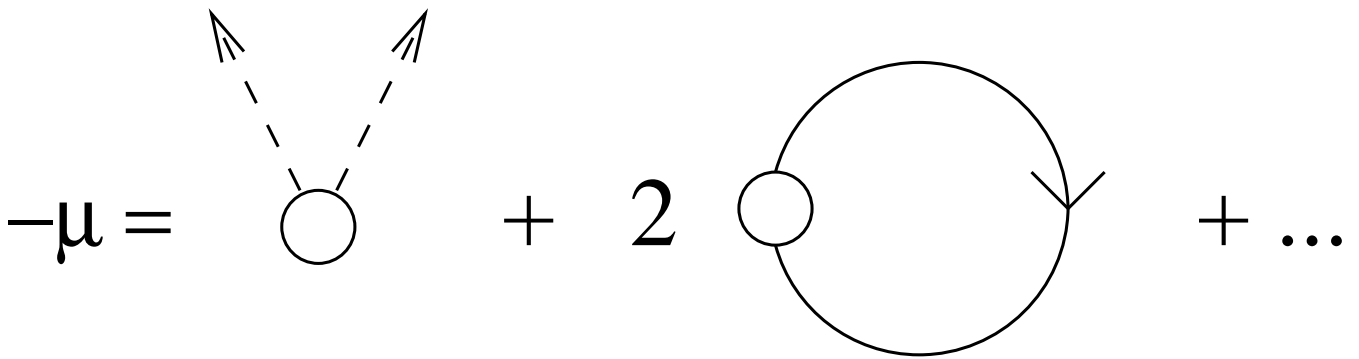}
\caption{\label{finalmufig}The chemical potential to one-loop order
after the ladder summation has been performed and the resulting
t-matrix been replaced by its expression in the dilute limit,
the constant $g$.}
\end{figure}

We note that no ladder counterparts to the
diagrams a and b in Fig.\ \ref{mufig} appear explicitly
in the expansion of the chemical
potential, since such diagrams are two-particle
reducible and are by construction excluded from the effective action 
$\Gamma$.
However, diagram b contains implicitly the ladder
contribution to diagram a. In order to establish this we
first simplify the notation by denoting
by $N_p$ the numerator of the exact normal Green's function
$G_{11}(p)$, which according to Eq.\ (\ref{momentumgreens}) is
$N_p = \omega+\epo-\mu+\Sigma_{11}(-p)$.
We then have $D_p = N_p N_{-p}
-\Sigma_{12}(p) \Sigma_{21}(p) =D_{-p} $, 
and the contribution from diagram b
can be rewritten on the forms
\begin{eqnarray}
\int d^4p U_{\pp} G_{12}(p) &=& \int d^4p U_{\pp}
\frac{\Sigma_{12}(p)}{N_pN_{-p}-\Sigma_{12}(p)\Sigma_{21}(p)}
\nonumber\\
&=& \int d^4p U_{\pp}\left(\frac{\Sigma_{12}(p)N_pN_{-p}}{D_p^2} -
\frac{\Sigma_{12}(p)\Sigma_{21}(p)\Sigma_{12}(p)}{D_p^2}\right)
\nonumber\\
&=& \int d^4p U_{\pp}\left(\Sigma_{12}(p) G_{11}(p) G_{11}(-p)
-\Sigma_{21}(p) G_{12}(p)^2\right)
\nonumber\\
&=& \int d^4p U_{\pp}\left(n_0 U_{\pp} G_{11}(p) G_{11}(-p) +
[\Sigma_{12}(p)-n_0U_{\pp}] G_{11}(p) G_{11}(-p)
-\Sigma_{21}(p) G_{12}(p)^2\right).
\label{laddermanip}
\end{eqnarray}
In Fig.\ \ref{laddermanipfig} the last two rewritings are depicted
diagrammatically.
We see immediately that
the first term on the right-hand side
corresponds to the first ladder contribution
to diagram a, and since to one-loop order, $\Sigma_{12}(p) = n_0 U_{\pp}$,
the other terms in Eq.\ (\ref{laddermanip}) are of two- and
higher-loop order. The self-energy in the
second term on the right-hand side can be
expanded to second loop order, and by iteration this yields all
the ladder terms, and the remainder can be kept track of analogously
to the way in which it is done in Eq.\ (\ref{laddermanip}). 
The resulting ladder resummed
diagrammatic expression for the chemical potential, displayed in
Fig.\ \ref{finalmufig}, is seen to be equal to that found in Ref.\ 
\cite{griffinreview}.

\begin{figure*}
\includegraphics[width=\textwidth]{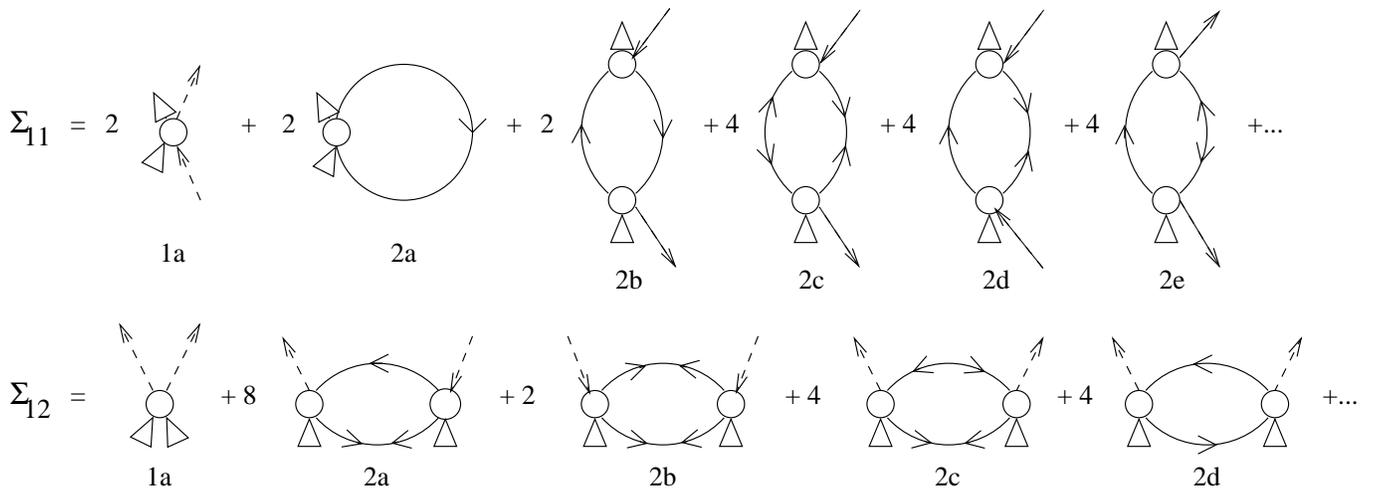}
\caption{\label{selffig}Normal, $\Sigma_{11}$, and anomalous,
$\Sigma_{12}$, self-energies to two-loop order after the ladder
summation has been performed.}
\end{figure*}

In the same manner, the self-energies are resummed. For
$\Sigma_{11}$ a straightforward ladder resummation of all terms is
possible, while for $\Sigma_{12}$ the same procedure as the one used
for diagrams a and b in Fig.\ \ref{mufig} for the chemical potential
has to be performed.
In Fig.\ \ref{selffig}, we show the resulting ladder resummed diagrams for
the self-energies $\Sigma_{11}$ and $\Sigma_{12}$ to two-loop order
in the dilute limit where $T(p,p',q) \approx g$.

In Ref.\ \cite{popov} a diagrammatic
expansion in the potential was performed, which yields to first order
the diagram $\Sigma^{(2a)}_{11}$ in Fig.\ \ref{selffig},
but not the other two-loop
diagrams. This theory, where the normal self-energy is taken to be
$\Sigma_{11}=\Sigma_{11}^{(1a)}+\Sigma_{11}^{(2a)}$, the anomalous
self-energy to $\Sigma_{12}=\Sigma_{12}^{(1a)}$, and the diagrams
displayed in Fig.\ \ref{finalmufig} are kept in the expansion of the
chemical potential, is referred to as the Popov approximation.
Although we showed at the end of Sec.\ \ref{homogensec}
that all the two-loop diagrams of Fig.\ \ref{selffig}
are of the same order of magnitude in the diluteness parameter
$\sqrt{n_0a^3}$ at zero temperature, the Popov
approximation applied at finite temperatures is justified, when the
temperature is high enough, $kT \gg g n_0$.
Below, we shall investigate the limits of validity at zero temperature
of the Popov
approximation in the trapped case.

In this and the preceding section we have shown how the well-known
expressions for
the self-energies and chemical potential for a homogeneous
dilute Bose gas
are conveniently obtained using the effective action formalism, where
they simply correspond to working to a particular order in the loop
expansion of the  effective action.  We have established that an
expansion in the diluteness parameter is equivalent to an expansion of
the effective action in
the number of loops.
 Furthermore, the method
provided a way of performing a systematic expansion,
and the results are easily generalized to finite temperatures.
We now turn to show that the effective action approach provides
a way of performing a systematic expansion even in
 the case of an inhomogeneous Bose gas.


\section{Inhomogeneous Bose gas}
\label{trappedsec}

We now review the experimentally relevant case of a Bose gas
trapped in an external static potential, in order to set the stage for the 
numerical calculations in Section \ref{numericsec}.
In this case the Bose gas will be spatially inhomogeneous.
The effective action formalism is equally capable of dealing with the
inhomogeneous gas, in which case all quantities are conveniently expressed
in configuration space as presented in section
\ref{formalismsec} . Varying, in accordance with Eq.\
(\ref{eq:motion1}),
the effective action $\Gamma$, Eq.\ (\ref{effaction}), with respect to
$\Phi^*(\rr,t)$, we obtain the equation of motion for the condensate
wave function:
\begin{equation}
(\ihdt-H_0+\mu)\Phi(\rr,t) =g \left|\Phi(\rr,t)\right|^2\Phi(\rr,t)
+ 2i g G_{11}(\rr,t,\rr,t) \Phi(\rr,t) -
\varder{\bar{\Gamma}_2}{\Phi^*(\rr,t)}.
\label{gpe}
\end{equation}
To zero-loop order, where only the first term
on the right-hand side appears, the equation
is the time-dependent Gross-Pitaevskii equation
\cite{gross,pitaevskii}.
We have already, as elaborated in the previous section,
performed the ladder summation by which the potential is renormalized
and the t-matrix appears and substituted its lowest order
approximation in the diluteness parameter, the constant $g$. Since
the t-matrix
in the momentum variables is a constant in the dilute limit, it
becomes in configuration space a product of three delta
functions,
$T(\rr_1,\rr_2,\rr_3,\rr_4)= g \delta (\rr_1-\rr_4)
\delta (\rr_2-\rr_4) \delta (\rr_3-\rr_4)$. The quantity
$\bar{\Gamma}_2$ is defined as the effective action obtained
from $\Gamma_2$ by summing the ladder terms whereby $U$ is replaced by
the t-matrix,
and its diagrammatic expansion is topologically of two-loop and higher order.

The Dyson-Beliaev equation, Eq.\ (\ref{dysonbel}), and the
equation determining the condensate wave function, Eq.\ (\ref{gpe}),
form a set of coupled
integro-differential self-consistency equations for the condensate
wave function and the Green's function, with the self-energy
specified in terms of the Green's function through
Eq.\ (\ref{sigmafraGamma}).
The Green's function can be conveniently
expanded in the amplitudes of the elementary excitations. We write the
Dyson-Beliaev equation, Eq.\ (\ref{dysonbel}), on the form
\begin{equation}\label{dyson2}
\int d\rr''dt'' \left[\ihdt\sigma_3\delta(\rr-\rr'')\delta(t-t'') +
\sigma_3L(\rr,t,\rr'',t'') \right]G(\rr'',t'',\rr',t')
= \hbar 1\delta(\rr-\rr')\delta(t-t'), \end{equation}
where we have introduced the matrix operator
$L(\rr,t,\rr',t')=\sigma_3H_0\delta(\rr-\rr')\delta(t-t')+
\sigma_3\Sigma(\rr,t,\rr',t')$ and $\sigma_3$ is a Pauli
matrix.
Up to one-loop order, the matrix $\Sigma$ is diagonal in the time and
space coordinates and
we can factor out the delta functions and write
$L(\rr,t,\rr',t')=\delta(t-t')\delta(\rr-\rr')L(\rr)$, where
\begin{equation}
L(\rr) = \left(\begin{array}{ll}
H_0 -\mu+ 2g|\Phi(\rr)|^2 & g\Phi(\rr)^2\\
-g\Phi^*(\rr)^2 & -H_0+\mu - 2g|\Phi(\rr)|^2
\end{array}\right).
\label{oneloopl}
\end{equation}
The eigenvalue equation for $L$ are the Bogoliubov
equations \cite{bogoliubov}. The Bogoliubov operator
$L$ is not Hermitian, but the operator
$\sigma_3 L$ is, which renders the eigenvectors of $L$ the
following properties \cite{fetter72,blaizot,castindum}.
For each eigenvector $\varphi_j(\rr) =
\left(u_j(\rr),v_j(\rr) \right)$
of $L$ with
eigenvalue $E_j$, there exists an eigenvector
$\tph_j(\rr)=(v_j^*(\rr),u_j^*(\rr))$ with eigenvalue
$-E_j$. Assuming the Bose gas
is in its ground state, the normalization
of the positive-eigenvalue eigenvectors can be chosen to be
$
\langle
\varphi_j,
\varphi_k
\rangle
=
\delta_{jk}
$
where we have introduced the inner product
\begin{equation}
\label{norm}
\langle
\varphi_j,
\varphi_k
\rangle
=
\int d\rr \varphi_j^{\dagger}(\rr) \sigma_3\varphi_k(\rr) =
\int d\rr ( u_j^*(\rr)u_k(\rr) - v^*_j(\rr)v_k(\rr)).
\end{equation}
It follows that the inner product of the
negative-eigenvalue eigenvectors $\tph$ are
\begin{equation}
\label{}
\langle
\tph_j,
\tph_k
\rangle
=
\int d\rr \tph_j^{\dagger}(\rr) \sigma_3\tph_k(\rr) =
\int d\rr ( v_j(\rr)v_k^*(\rr) - u_j(\rr)u_k^*(\rr)) = -\delta_{jk}
\end{equation}
and
the eigenvectors $\varphi$ and $\tph$ are mutually orthogonal,
$\langle \varphi_j, \tph_k \rangle = 0$.
By virtue of the Gross-Pitaevskii equation, the
vector $\varphi_0(\rr) =(\Phi(\rr),-\Phi^*(\rr))$ is an eigenvector
of the Bogoliubov operator $L$  with zero
eigenvalue and zero norm. In order to obtain a completeness relation,
we must also introduce the vector
$\varphi_a(\rr)=(\Phi_a(\rr),-\Phi^*_a(\rr))$ satisfying the relation
$
L\varphi_a = \alpha \varphi_0,
$
where $\alpha$ is a constant determined by normalization,
$\langle
\varphi_0,
\varphi_a
\rangle
= 1.$
The resolution of the identity then becomes
\begin{equation}
{\sum_{j}}'\left(\varphi_j(\rr)\varphi_j^{\dagger}(\rr') - \tph_j(\rr)
\tph_j^{\dagger}(\rr')\right)\sigma_3 +
\left(\varphi_a(\rr)\varphi_0^{\dagger}(\rr')+
\varphi_0(\rr)\varphi_a^{\dagger}(\rr')\right)
\sigma_3 = 1 \delta(\rr-\rr'),
\label{completeness01}
\end{equation}
where the prime on the summation sign indicates that
the zero-eigenvalue mode
$\varphi_0$ is excluded from the sum.
Using the resolution of the identity, Eq.\ (\ref{completeness01}),
allows us to invert Eq.\ (\ref{dyson2}) to obtain the well-known spectral 
representation of the Green's function
\begin{equation}\label{spectralgreen}
G(\rr,\rr',\omega) =
\hbar{\sum_j}'
\left(
\frac{1}{-\hbar\omega +E_j}\varphi_j(\rr)\varphi^{\dagger}_j(\rr')
-\frac{1}{-\hbar\omega -E_j}\tph_j(\rr)\tph^{\dagger}_j(\rr')
\right).
\end{equation}
It follows from the spectral representation of the Green's function,
that the
eigenvalues $E_j$ are the elementary
excitation energies of the condensed gas (here constructed explicitly
to one-loop order). Using Eq.\ (\ref{spectralgreen}), we can 
at zero temperature express the noncondensate density or the depletion of the
condensate,
$\nnc=n-n_0$, in terms of the Bogoliubov amplitudes
\begin{equation}
\nnc(\rr) = i \int \frac{d\omega}{2 \pi} G_{11}(\rr,\rr,\omega)
= {\sum_j}' |v_j(\rr)|^2.
\label{noncondensate}
\end{equation}


\section{loop expansion for a trapped bose gas}
\label{numericsec}

We now turn to determine the validity criteria for the  equations
obtained to various orders in the loop
expansion for the ground state of a Bose gas trapped in an isotropic
harmonic potential $V(r)=\frac12m\omega_{\rm t}^2r^2$. To this end, we
shall numerically compute the self-energy diagrams to different
orders in the loop expansion.

Working consistently to one-loop order, we need only employ  
Eq.\ (\ref{gpe}) to zero loop order, providing
the condensate wave function which upon insertion into
Eq.\ (\ref{oneloopl}) yields
the Bogoliubov operator $L$ to one-loop order, from which the Green's function
to one-loop order is obtained from Eq.\ (\ref{spectralgreen}).
The resulting Green's function
is then used to calculate the various self-energy terms numerically.
In order to do so, we
make the equations dimensionless with the transformations
$r = \aos \tilde{r}$,
$\Phi = \sqrt{N_0/\aos^3} \tilde{\Phi}$,
$u_j = \aos^{-3/2} \tilde{u}_j$,
$E_j = \hbar \omega_{\rm t}\tilde{E}_j$, and
$g = (\hbar\omega_{\rm t}\aos^3/N_0) \tilde{g}$,
where $\aos = \sqrt{\hbar/m\omega_{\rm t} }$ is the oscillator length
of the harmonic trap, and $N_0$
is the number of bosons in the condensate.

To zero-loop order, the time-independent
Gross-Pitaevskii equation on dimensionless
form reads
\begin{equation}\label{dimlessgpe}
 - \frac12\nabla^2_{\tilde{r}}\tilde{\Phi} +
  \frac12 \tilde{r}^2\tilde{\Phi} + \bg|\tilde{\Phi}|^2\tilde{\Phi}
  = \tilde{\mu}\tilde{\Phi}.
\end{equation}
We solve Eq.\ (\ref{dimlessgpe}) numerically with the
steepest-descent method, which has proven to be sufficient
for solving the present equation \cite{steepest}.
The result thus obtained for $\tilde{\Phi}$ is inserted into the
one-loop expression for the Bogoliubov operator $L$,
Eq.\ (\ref{oneloopl}),
in order to calculate the Bogoliubov amplitudes $\tilde{u}_j$ and
$\tilde{v}_j$ and the eigenenergies $\tilde{E}_j$. Since the
condensate wave function for the ground-state,
$\tilde{\Phi}$, is real and rotationally symmetric,
the amplitudes $\tilde{u}_j, \tilde{v}_j$ in the Bogoliubov
equations can be labeled by the two angular momentum quantum
numbers $l$ and $m$, and a radial quantum number $n$, and we write
$\tilde{u}_{nlm}(\tdr, \theta, \phi)=
\tilde{u}_{nl}(\tdr)Y_{lm}(\theta, \phi)$,
$\tilde{v}_{nlm}(\tdr, \theta, \phi)=
\tilde{v}_{nl}(\tdr)Y_{lm}(\theta, \phi)$.
The resulting Bogoliubov equations are
linear and one-dimensional:
\begin{eqnarray}
  \left(-\frac12\frac{1}{\tdr}\frac{\partial^2}{\partial \tdr^2}\tdr
  + \frac12 \frac{l(l+1)}{\tdr^2} +
  \frac12\tdr^2-\tilde{\mu}
  +2\bg\tilde{\Phi}^2(\tdr)\right) \tilde{u}_{nl}(\tdr) +
  \bg\tilde{\Phi}^2(\tdr) \tilde{v}_{nl}(\tdr) = \tilde{E}_{nl}
  \tilde{u}_{nl}(\tdr),
   \nonumber\\
   \label{dimlessbe}
  \left(-\frac12\frac{1}{\tdr}\frac{\partial^2}{\partial \tdr^2}\tdr
  + \frac12 \frac{l(l+1)}{\tdr^2} +
  \frac12\tdr^2-\tilde{\mu}
  +2\bg\tilde{\Phi}^2(\tdr)\right) \tilde{v}_{nl}(\tdr) +
  \bg\tilde{\Phi}^2(\tdr) \tilde{u}_{nl}(\tdr) = -\tilde{E}_{nl}
  \tilde{v}_{nl}(\tdr).
\end{eqnarray}
Note that the only parameter in the problem is the dimensionless
coupling parameter $\bg=4\pi N_0a/\aos$. Solving the Bogoliubov equations
reduces to diagonalizing  the band diagonal $2M\times 2M$
matrix $L$, where $M$ is the size of the numerical
grid. The value of $M$ in our computations was varied between 180
and 360, higher values for stronger coupling, and the grid
constant has been chosen to $0.05\aos$ giving a maximum system size of
$18\aos$. We have used MATLAB to perform
the diagonalization.

In the following we shall estimate
the orders of magnitude and the parameter dependence of the different
two- and three-loop self-energy diagrams, and to this end we shall use
the one-loop results for the amplitudes
$\tilde{u}$, $\tilde{v}$ and the eigenenergies $\tilde{E}$ obtained
numerically.
When working to two- and three-loop order, one must also consider
the corresponding corrections to the approximate
t-matrix $g$. These contributions
have been studied in Ref.\ \cite{proukakis2}, and their inclusion
will not lead to any qualitative changes of our results. In fact, even 
at finite temperature, the dependence of the t-matrix on the coupling 
is weak as long as the temperature is not close to the critical 
temperature for Bose-Einstein condensation $T_c$ 
\cite{griffinreview,stoof}.

\begin{figure}[htbp]
\includegraphics[width=\columnwidth]{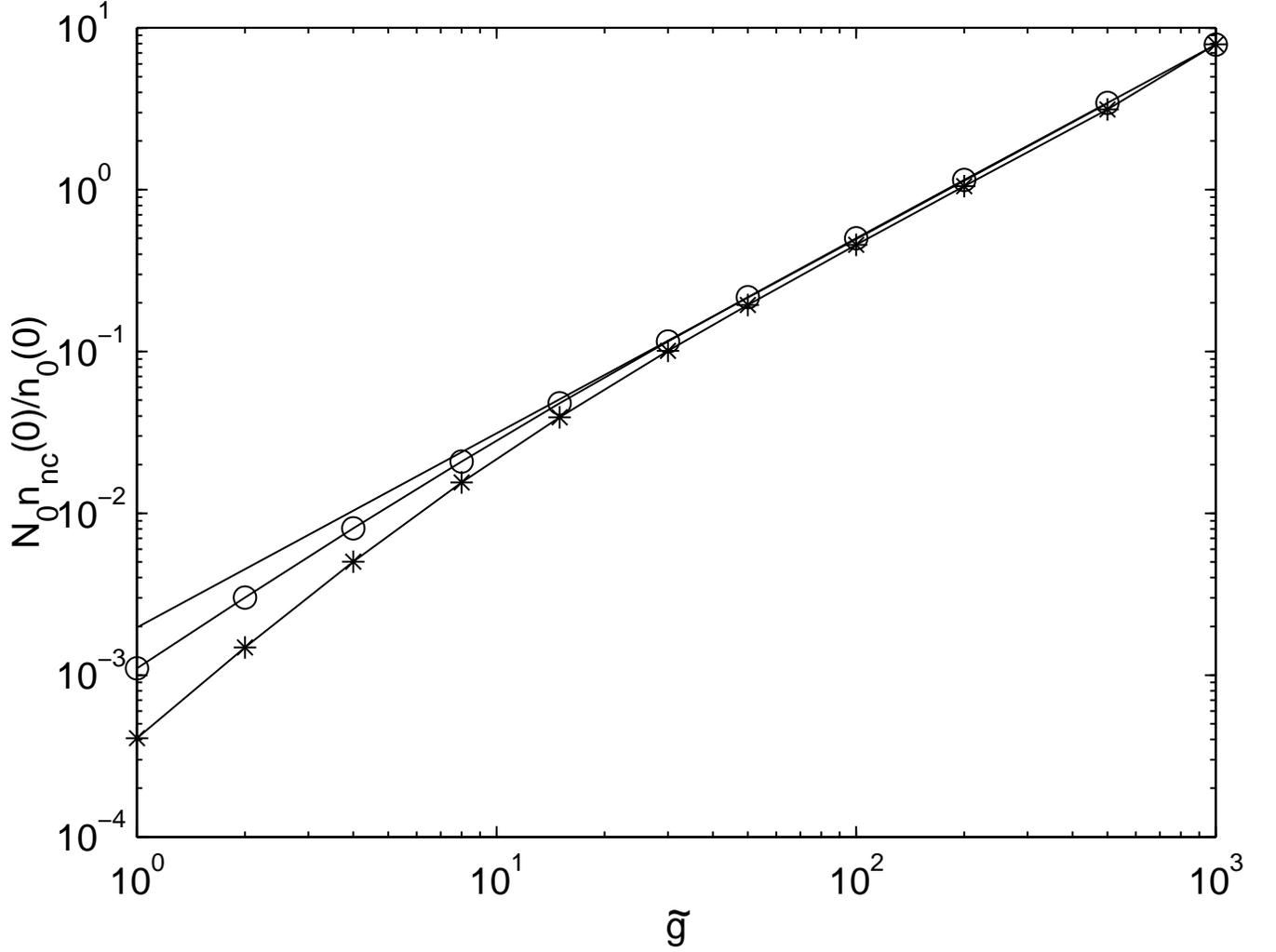}
\caption{\label{depletionfig}Fractional depletion of the condensate
at the trap center
as a function of the dimensionless coupling strength
$\bg=4\pi N_0a/\aos$. Asterisks denote our numerical results, circles
denote the local-density approximation with the numerically computed
condensate density inserted, and the line is the local-density
approximation using the Thomas-Fermi approximation for the condensate
density.}
\end{figure}

Let us first compare the one-loop and two-loop contributions to the normal
self-energy.
The only one-loop term is
\begin{equation}
\Sigma_{11}^{(1a)}(\rr,\rr',\omega) = 2 g |\Phi(\rr)|^2 \delta(\rr-\rr')
= 2gn_0(\rr) \delta(\rr-\rr').
\end{equation}
We first compare $\Sigma_{11}^{(1a)}$ with the
two-loop term which is proportional to a delta function, i.e.,
the diagram 2a in
Fig.\ \ref{selffig}. We shall shortly compare this diagram to
the other two-loop diagrams. For diagram 2a we have
\begin{equation}
\Sigma_{11}^{(2a)}(\rr,\rr',\omega) = 2ig \delta(\rr-\rr')
\int \frac{d\omega'}{2\pi} G(\rr,\rr,\omega')
= 2g\nnc (\rr) \delta(\rr-\rr').
\end{equation}
The ratio of the two-loop to one-loop self-energy contributions at the
point $\rr$ is thus equal to the fractional depletion of the
condensate at that point. In Figure \ref{depletionfig} is
shown the numerically computed dimensionless fractional
depletion at the origin, $\bnc(0)/\tilde{n}_0(0)$,
where we have introduced the dimensionless notation
\begin{eqnarray}
\tilde{n}_0(\tilde{\rr}) = |\tilde{\Phi}(\tilde{\rr})|^2,
\nonumber\\
\bnc(\tilde{\rr}) = {\sum_j}' |\tilde{v}_j(\tilde{\rr})|^2.
\end{eqnarray}
We have chosen to evaluate the densities at the origin, $\rr=0$, in order to
avoid a prohibitively large summation over the $l \neq 0$
eigenvectors.
As apparent from Figure \ref{depletionfig}, the log-log
curve has a slight bend initially, but becomes almost straight for
coupling strengths $\bg\agt 1000$.
A logarithmic fit to the straight portion of the curve
gives the relation
\begin{equation}
\frac{
\bnc(0)
}{
\tilde{n}_0(0)
} \simeq
0.0019\bg^{1.2}.
\label{fracdepletion2001}
\end{equation}
For weaker coupling, the curvature is seen to be stronger.
When we reintroduce dimensions, the power-law relationship, 
Eq.\ (\ref{fracdepletion2001}),
is multiplied by the inverse of the number of bosons in the
condensate, $N_0^{-1}$,
because the actual and dimensionless self-energies
are related according to
\begin{equation}
\Sigma^{(s)} = \frac{\hbar\omega_{\rm t}
\aos^3}{N_0^{s-1}}\tilde{\Sigma}^{(s)},
\label{sigmadimens}
\end{equation}
where $s$ denotes the loop order in question.
The ratio between different
loop orders of the self-energy is thus not determined solely by the
dimensionless coupling parameter $\bg=4\pi N_0a/\aos$, but
by $N_0$ and $a/\aos$ separately. We thus obtain for the fractional
depletion in the strong-coupling limit, $\bg\agt 1000$,
\begin{equation}
\frac{\nnc(0)}{n_0(0)} = \frac1{N_0}
\frac{\bnc(0)}{\tilde{n}_0(0)} \approx
0.041 N_0^{0.2}\left(\frac{a}{\aos}\right)^{1.2}.
\label{numdepletion}
\end{equation}

It is of interest to compare our numerical results with approximate
analytical results such as those
obtained using the
local density approximation (LDA). The LDA amounts to substituting a
coordinate-dependent condensate density in the expressions
valid for the
homogeneous gas. The homogeneous-gas result for
the fractional depletion is \cite{bogoliubov}
\begin{equation}
\frac{\nnc}{n_0} = \frac{8}{3\sqrt{\pi}}\sqrt{n_0a^3}.
\label{lda}
\end{equation}
In the strong coupling limit we can use the
Thomas-Fermi approximation for the condensate density
\begin{equation}
n_0(\rr) = \frac{1}{8\pi \aos^2 a}\left(\frac{15N_0a}{\aos}\right)^{2/5}
\left[1-\left(\frac{\aos}{15N_0a}\right)^{2/5} \frac{r^2}{\aos^2}\right],
\end{equation}
which is obtained by neglecting the kinetic term in the
Gross-Pitaevskii equation \cite{bp}. For the fractional depletion at the
origin there results in the
local density approximation
\begin{equation}
\frac{\nnc(0)}{n_0(0)} = \frac{(15N_0)^{1/5}}{3\pi^2\sqrt{2}}
\left(\frac{a}{\aos}\right)^{6/5},
\label{rmpdepletion}
\end{equation}
as first obtained in Ref.\ \cite{rmp}.
The LDA is a valid approximation when the gas
locally resembles that of a homogeneous system, i.e.,
when the condensate wave function changes little on the scale of the coherence
length $\xi$, which according to the Gross-Pitaevskii equation is
$\xi = (8\pi n_0(0)a)^{-1/2}$.
For a trapped cloud of bosons in the ground state, its radius $R$
determines the rate of change of the density profile.
Since $R$ is a factor $\bg^{2/5}$ larger than
$\xi$ \cite{bp}, we expect the agreement between the LDA and the exact
results to be best in the strong-coupling regime.
The fractional depletion of the condensate
at the trap center
as a function of the dimensionless coupling strength
$\bg=4\pi N_0a/\aos$ is shown in Fig.\ \ref{depletionfig}.
In Fig.\ \ref{depletionfig} are displayed both the local-density result
Eq.\ (\ref{lda}) with the numerically computed condensate density
inserted, and the Thomas-Fermi approximation (\ref{rmpdepletion}), 
showing that
the LDA indeed is valid when the coupling is strong. Furthermore,
inspection of Eq.\ (\ref{rmpdepletion}) reveals that the LDA
coefficient and exponent agree with the numerically found result of
Eq.\ (\ref{numdepletion}), which is valid for strong coupling.
However, when $\bg \alt 10$, the LDA prediction for the depletion
deviates significantly from the
numerically computed depletion. Inserting the
numerically obtained condensate density into the LDA instead of the
Thomas-Fermi approximation is seen not to substantially improve the result.

The relation for the fractional depletion, Eq.\  (\ref{numdepletion}), 
is in agreement with the results of
Ref.\ \cite{stenholm}, where the leading-order corrections to the
Gross-Pitaevskii equation were considered in the
one-particle irreducible effective-action formalism,
employing physical assumptions about the relevant length scales
in the problem. These leading-order corrections 
were found to have the same
power-law dependence on $N_0$ and $a/\aos$.
A direct comparison of the
prefactors cannot be made, because the objective of
Ref.\ \cite{stenholm} was to estimate
the higher-loop correction terms to the Gross-Pitaevskii equation and
not to the self-energy.

The two-loop term $\Sigma_{11}^{(2a)}$ can at zero temperature
be ignored as long as $\bnc \ll \tilde{n}_0$,
which is true in a wide, experimentally relevant parameter regime.
The result for the fractional depletion Eq.\
(\ref{numdepletion}) depends very weakly on $N_0$, so as long as $N_0$
does not exceed $10^9$, which is usually fulfilled in experiments, we
can restate the criterion for the validity of Eq.\ (\ref{numdepletion})
into the condition $a \ll
\aos$. In experiments on atomic rubidium and sodium condensates,
this condition is fulfilled, except in the instances where Feshbach
resonances are used to enhance the scattering length
\cite{feshbach,jilafeshbach}.

\begin{figure}
\includegraphics[width=\columnwidth]{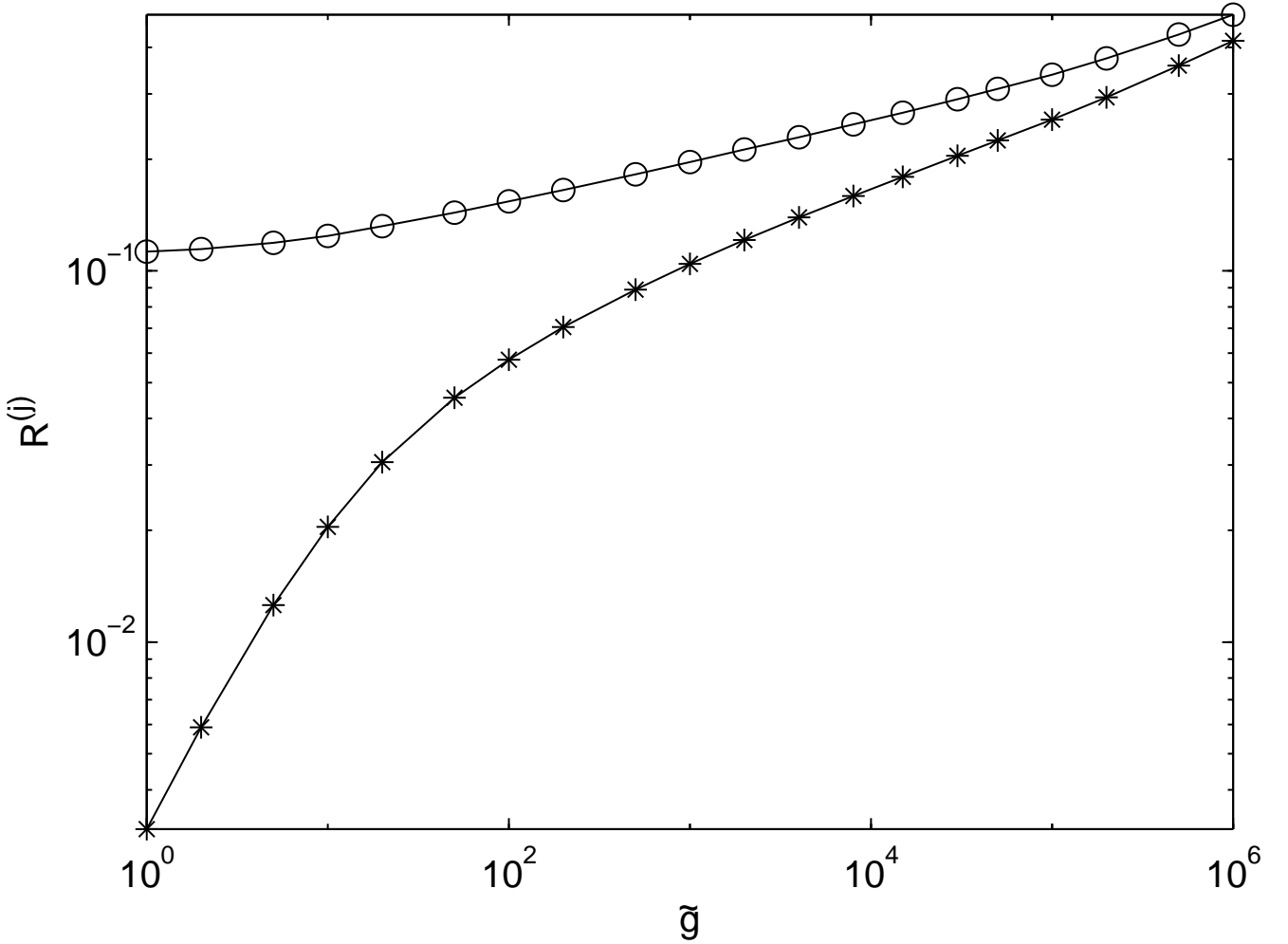}
\caption{\label{sigma2fig}Ratio between different dimensionless two-loop
self-energy terms as functions of the dimensionless coupling strength
$\bg=4\pi N_0a/\aos$.
Asterisks denote the ratio $R^{(2b)}$ as defined in Eq.\
(\ref{twoloopratio}) and circles denote the
ratio $R^{(2c)}$. The terms $R^{(2d)}$ and $R^{(2e)}$
are equal and turn out to be similar in magnitude to $R^{(2b)}$,
and are not displayed.}
\end{figure}

In section \ref{homogensec} we showed that for a homogeneous gas
all two-loop diagrams are
equally important in the sense that they are all of the same order in
the diluteness parameter $\sqrt{n_0a^3}$. The situation in a
trapped system is not so clear, since the density is not constant.
We shall therefore compare the five
normal self-energy diagrams $\Sigma_{11}^{(2a-e)}$ in  Fig.\
\ref{selffig},
to see whether they
display the same parameter dependence and whether any of the terms can
be neglected. In particular, the Popov approximation
contains the diagram $\Sigma_{11}^{(2a)}$ but neglects
all other second-order diagrams, and we will now determine its limits
of validity at zero temperature.
Since diagram 2a contains a delta function, we shall
integrate over one of the spatial arguments of the self-energy
terms and keep the other one fixed at the origin, $\rr=0$.
We denote by $R^{(j)}$ the ratio between the integrated self-energy
terms j and 2a,
\begin{equation}
R^{(j)} =\frac{\int d\rr \Sigma_{11}^{(j)}(0,\rr,\omega=0)}
{\int d\rr \Sigma_{11}^{(2a)}(0,\rr,\omega=0)}.
\label{twoloopratio}
\end{equation}
In Fig.\ \ref{sigma2fig}, we display the ratios $R^{(j)}$ for
the different integrated self-energy contributions corresponding to
the diagrams where j represents 2b and 2c. The contributions from diagrams 
2d and 2e are equal and turn out to be exactly equal within our numerical 
precision to that from diagram 2c. Furthermore, inspection of the diagrams 
in Fig.\ \ref{selffig} reveals that when the condensate wave function is 
real, the anomalous contribution 
$\Sigma_{12}^{(2a)}$ is equal to $\Sigma_{11}^{(2d)}$; 
$\Sigma_{12}^{(2b)}$ and $\Sigma_{12}^{(2c)}$ are equal to 
$\Sigma_{11}^{(2c)}$, and $\Sigma_{12}^{(2d)}$ is equal to 
$\Sigma_{11}^{(2b)}$.
In the parameter regime displayed here, the contribution from
diagram 2a is larger than the others by approximately a factor of
ten, and displays only
a weak dependence on the coupling strength. In the weak-coupling
limit, $\tilde{g} \alt 1$, it is seen that
the terms corresponding to diagrams 2b-e can be neglected as
in the Popov approximation, with an
error in the self-energy of a few per cent. When the coupling gets
stronger, this correction becomes more important. A power-law fit
to the ratio $R^{(2b)}$ in the regime
where the log-log curve is straight yields the dependence
\begin{equation}
R^{(2b)} \approx 0.028 \bg^{0.19},
\end{equation}
which is equal to 0.5 when $\bg \approx 10^6$; for $\bg$ greater
than this value, the Popov approximation is seen not to be valid.
If the ratio between the oscillator length and the
scattering length is equal to one hundred, $\aos= 100 a$,
the Popov approximation deviates
markedly from the two-loop result
when $N_0$ exceeds $10^7$, which is often the case experimentally.

\begin{figure}
\includegraphics[width=\columnwidth]{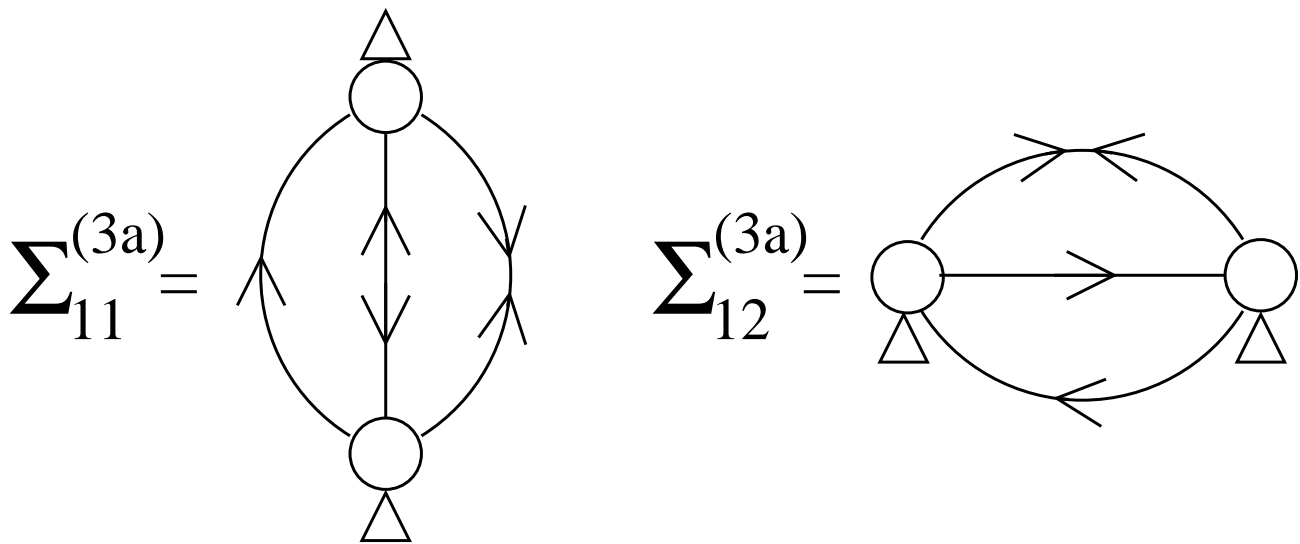}
\caption{\label{diag3a}Self-energy diagrams to three loop order which
are evaluated numerically.}
\end{figure}

\begin{figure}
\includegraphics[width=\columnwidth]{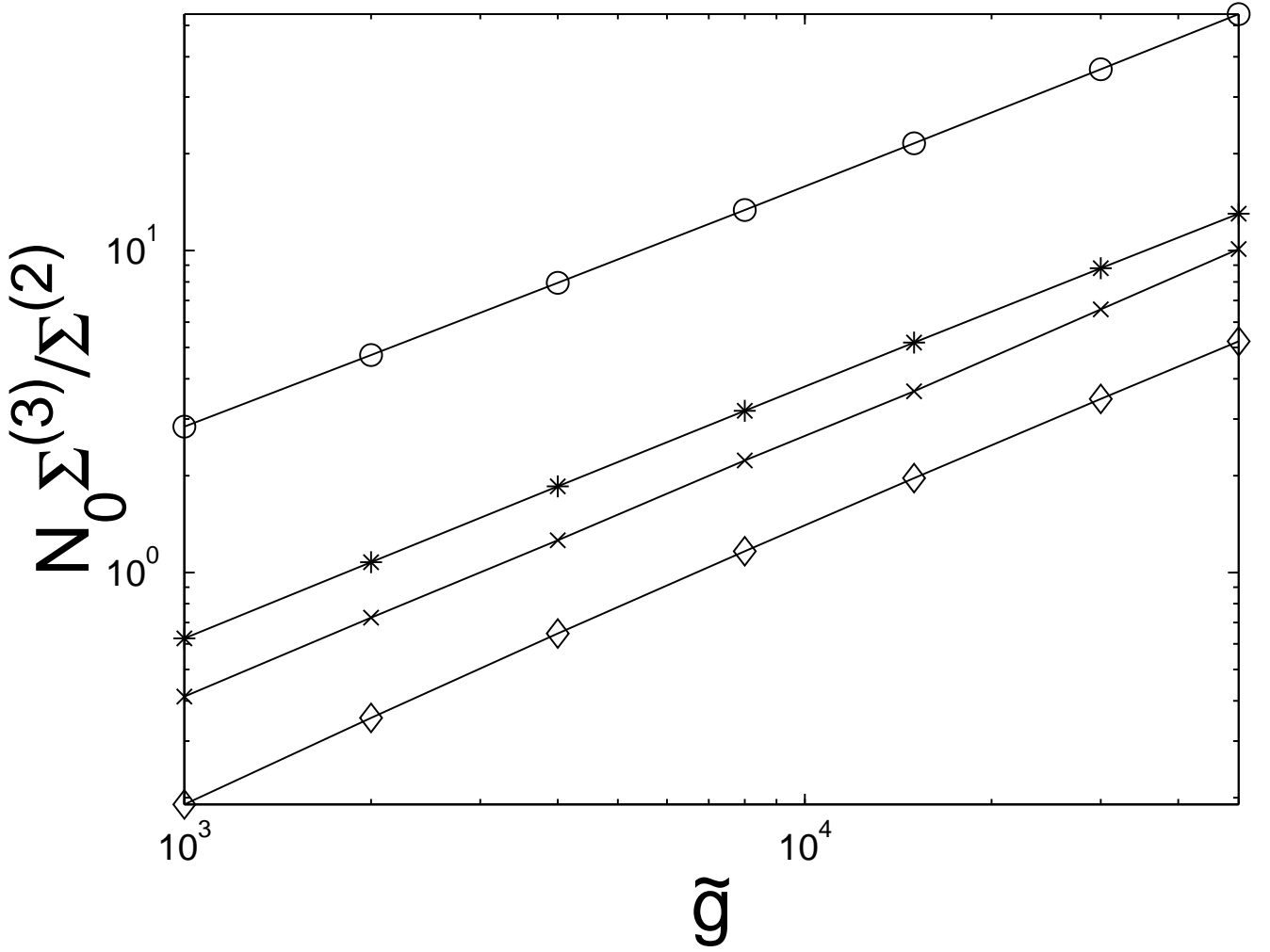}
\caption{\label{sigma32fig}Ratio of dimensionless three-loop to 
two-loop
self-energy diagrams as a function of the dimensionless coupling
strength $\bg=4\pi N_0a/\aos$. Asterisks denote the ratio of the normal
self-energy terms
$N_0\Sigma_{11}^{(3a)}/\Sigma_{11}^{(2b)}$ evaluated at
the point $(0,\aos,\omega=0)$, open circles denote the same ratio
evaluated at $(0,0.5\aos,\omega=0)$, and diamonds denote the same ratio
evaluated at $(0,1.5\aos,\omega=0)$.
Crosses denote the ratio
of anomalous self-energy terms
$N_0\Sigma_{12}^{(3a)}/\Sigma_{12}^{(2a)}$ at
$(0,\aos,\omega=0)$.}
\end{figure}

In order to investigate the importance of higher-order terms in the
loop expansion, we proceed to study the three-loop self-energy
diagrams. We have found the number of summations over Bogoliubov
levels to be prohibitively large for most three-loop terms;
however, we {\it have} been able to compute the two diagrams
$\Sigma_{11}^{(3a)}$ and $\Sigma_{12}^{(3a)}$, displayed in Fig.\
\ref{diag3a}, for the case where one of the spatial arguments is
placed at the origin
thereby avoiding a summation over $l\neq 0$ components.
We compare the diagrams $\Sigma_{11}^{(3a)}$ and
$\Sigma_{12}^{(3a)}$ to the two-loop diagrams. As we have
seen, diagrams $\Sigma_{11}^{(2b)}$, $\Sigma_{11}^{(2c)}$ and
$\Sigma_{11}^{(2d)}$ in Fig.\ \ref{selffig} are similar in
magnitude and dependence on $\bg$, as are the anomalous two-loop
diagrams $\Sigma_{12}^{(2a-d)}$; we have therefore chosen to
evaluate only diagrams $\Sigma_{11}^{(2b)}$ and
$\Sigma_{12}^{(2a)}$. The results for the ratios
$\tilde{\Sigma}_{11}^{(3a)}(0,r,\omega=0)/
\tilde{\Sigma}_{11}^{(2b)}(0,r,\omega=0)$
and
$\tilde{\Sigma}_{12}^{(3a)}(0,r,\omega=0)/
\tilde{\Sigma}_{12}^{(2a)}(0,r,\omega=0)$,
evaluated for different choices of $r$,
are shown in Fig.\ \ref{sigma32fig}.
A linear fit to the log-log plot gives for the
normal terms the coefficient 0.016 and the exponent 0.76 when
$r=0.5\aos$ and the coefficient 0.0029 and the exponent 0.78 when
$r=\aos$, and for the anomalous terms with the choice $r=\aos$
the coefficient is 0.0015 and the exponent 0.82. Restoring
dimensions according to Eq.\ (\ref{sigmadimens}) we obtain
\begin{equation}
\label{3loopfit}
\frac{\Sigma_{11}^{(3a)}(0,\aos,\omega=0)}
{\Sigma_{11}^{(2b)}(0,\aos,\omega=0)} \approx
0.15 N_0^{-0.2} \left(\frac{a}{\aos}\right)^{0.8}.
\end{equation}
The ratio between three- and two-loop self-energy terms in the
homogeneous case was in Sec.\ \ref{homogensec} found to be
proportional to $\sqrt{n_0a^3}$. A straightforward application of
the LDA, inserting the central density $n_0(0)$ into this formula,
yields the dependence $\Sigma_{11}^{(3a)}/ \Sigma_{11}^{(2b)}
\propto N_0^{0.2}(a/\aos)^{1.2}$. This is not in accordance with the
numerical result, Eq.\ (\ref{3loopfit}), although the self-energies were
evaluated at spatial points close to the trap center. The discrepancy
between the LDA and the numerical result is attributed to the fact that
we fixed the spatial points in units of $\aos$ while varying the
coupling $\bg$, although the physical
situation at the point $r=\aos$ (and $r=\frac12 \aos$ and
$r=\frac32 \aos$
respectively) varies when $\bg$ is varied.
It is possible that the agreement with the LDA had been better 
if the length scales had been fixed in units of 
the actual cloud radius (as given by the Thomas-Fermi approximation) 
rather than the oscillator length. However, the present calculation 
agrees fairly well with the 
LDA as long as the number of atoms in the
condensate lies within reasonable bounds.
Since $N_0 > 1$ in the condensed state,  Eq.\ (\ref{3loopfit}) yields 
that $\Sigma_{11}^{(3a)} \ll \Sigma_{11}^{(2b)}$
whenever the s-wave scattering length is much smaller than the trap
length. We conclude that only when this condition is not
fulfilled is it necessary to study diagrams of three-loop order
and beyond.

\section{Conclusion}
\label{conclusionsec}
We have applied the two-particle irreducible
effective-action approach to a condensed Bose gas,
and shown that it allows
for a convenient and systematic derivation of the equations of
motion both in the homogeneous and trapped
case. We have chosen to work at zero temperature, but the formalism is
with equal ease capable of dealing with systems at finite temperatures
and general non-equilibrium states.
Beliaev's  diagrammatic expansion in the diluteness parameter
and the t-matrix equations are expediently arrived at
with the aid of the effective-action
formalism. We have shown that the parameter characterizing the loop expansion
for a homogeneous Bose gas
is equal to the diluteness parameter,
the ratio of the s-wave scattering length and the inter-particle
spacing.
For an isotropic, three-dimensional harmonic-oscillator trap at zero
temperature, the small parameter governing the loop expansion
has been found to depend on the ratio between the s-wave scattering
length and the oscillator length of the trapping potential, and
has a weak dependence on the number of particles.
The expansion to one-loop order, and hence the Bogoliubov equation,
is found to provide a valid description for the zero-temperature
trapped gas when the oscillator length
exceeds the s-wave scattering length. We have compared our numerical
results with the local-density approximation, which is found to be
valid when the number of particles is large compared to the
ratio between the oscillator length and 
the s-wave scattering length.
Furthermore, we have found that all the
self-energy terms of two-loop order are not equally large
for the case of a trapped
system: in the limit when the number of particles is not large 
compared to the ratio between the oscillator length and 
the s-wave scattering length, 
the Popov approximation has been shown to be a valid approximation.

Recent experiments have probed the regime of large scattering
lengths in rubidium and sodium condensates \ref{feshbach,jilafeshbach}.
In the limit of very strong interactions, the loop expansion itself is
questionable and other theoretical methods must be invoked
\cite{cowell}. However, in the intermediate regime of moderately dense
clouds, the self-energy corrections considered in this paper should
become important and are possible to study experimentally.

\begin{acknowledgments}
It is a pleasure to acknowledge helpful discussions with
Lars Melwyn Jensen, Chris Pethick, Henrik Smith, and
Stig Stenholm.
\end{acknowledgments}

\end{document}